\documentclass[prl,twocolumn,showpacs]{revtex4}
\usepackage{epsfig}
\usepackage{dcolumn}
\usepackage{graphicx}
\usepackage{subfigure}
\usepackage{epsf}
\usepackage{epstopdf}
\begin{document}

\title{Chain metallicity and antiferro-paramagnetism competition in underdoped YBa$_2$Cu$_3$O$_{6+x}$: a first principles description}
\author{Alessio Filippetti}
\author{Giorgia M. Lopez}
\author{Mauro Mantega}
\author{Vincenzo Fiorentini}
\affiliation{SLACS CNR-INFM and Dipartimento
di Fisica, Universit\`a di Cagliari, I-09042 Monserrato (CA), Italy}

\date{\today}

\begin{abstract}
We describe from advanced first principles calculations the energetics of oxygen doping  and its relation to insulator-metal transitions in underdoped YBa$_2$Cu$_3$O$_{6+x}$. We find a strong tendency of doping oxygens to order into non-magnetic Cu$^{1+}$O$_x$ chains at any $x$. Ordering produces one-dimensional metallic bands, while configurations with non-aligned oxygens are insulating. The  Cu$^{2+}$O$_2$ planes remain insulating and antiferromagnetic up to a threshold between $x$=0.25 and 0.5, above which a paramagnetic  normal-metal state prevails. The in-plane antiferro-paramagnetic competition depends on $x$, but only weakly on the ordering state of the chains.
\end{abstract}

\pacs{Valid Pacs}
\maketitle
A central puzzle in the physics of the archetypal high-T$_c$ superconducting cuprate YBa$_2$Cu$_3$O$_{6+x}$ is the occurrence of  several seemingly concurrent  phase transitions --structural, magnetic, order-disorder, insulating-metal, and superconducting--  in the underdoped region of the (T,$x$) phase diagram. Superconductivity sets on  at $x_c$$\sim$0.35, so that an insulator-metal  transition (IMT)   should occur   around the same doping. This IMT, probably associated with a change in magnetic state,  is thought be related to  an order-disorder transition in the Cu-O chains \cite{yu}; holes form a Fermi glass in the low-doping Mott insulator up to a  density h$\sim$0.07 per CuO$_2$ (roughly corresponding to $x_c$), above which they start percolating in the planes leading to a conventional metal state.  Experiments, however, suggest that {\it CuO chains} undergo an independent IMT at lower doping than the CuO$_2$ planes. At strong to  moderate underdoping ($x$$\sim$0.2-0.8), infrared experiments \cite{lee} identify the electromagnetic response of the chains  as that of a Tomonaga-Luttinger liquid-like one-dimensional conductor, although the chain fragments are too short ($\sim$15-400 \AA) to support dc currents across macroscopic regions (except for weak underdoping ($x$=0.95) whereby they are conducting \cite{yoichi,basov}). Optical data \cite{widder} show that  oxygens cluster in metallic islands with local orthorhombic symmetry even at minute doping, and that  metallic percolation through the chains occurs at a doping as low as $x$=0.1-0.2. 

It is fair to say that the relation of chain ordering with one-dimensional metallization and with the IMT in the CuO$_2$ planes at higher doping has never been described thoroughly from first principles. Here, using advanced first principles electronic structure calculations, we a) study the connection between intra-chain one-dimensional metallicity and oxygen ordering in (non-magnetic) CuO chains as function of doping, and b) analyze the competition of  the antiferromagnetic (AF) insulating and paramagnetic (PM) metallic states of the CuO$_2$ planes in different chain ordering configurations.  The picture we arrive at is as follows. YBa$_2$Cu$_3$O$_6$ is, as expected, a Mott-Hubbard insulator. At small $x$ it remains everywhere  insulating inasmuch as doping oxygens are disorderly distributed through the chains. A dispersed,  one-dimensional  metallic band appears when oxygens orderly align into CuO chains.  Chain alignment is found to be always favored,  consistently with the evidence that chain nucleation starts immediately at $x$$>$0 and that metallic percolation in the chains  is already detectable at $x$$\sim$0.1 \cite{lee,widder}. The dispersed band has a strongly dominant chain-atom character, and holes remain largely bound to the doping oxygen O(1) and the nearby Cu(1) ions. Only a minor hole fraction is transferred to the CuO$_2$ planes (indicating a rather sharp decoupling of the two playgrounds): the ensuing small  variations of orbital occupations  govern the  very tight competition between the in-plane insulating AF state and the conventional metallic PM phase. The AF state gives way to the PM  metallic state at a doping between $x$=0.25 and 0.5, consistent with the overall metallization threshold implied by the superconductive onset: intriguingly, this crossover occurs {\it independently} of the way O(1)'s are distributed in the chains, i.e. on the order-disorder interplay  governing the in-chain IMT.

\begin{figure}
\includegraphics[clip,width=8cm]{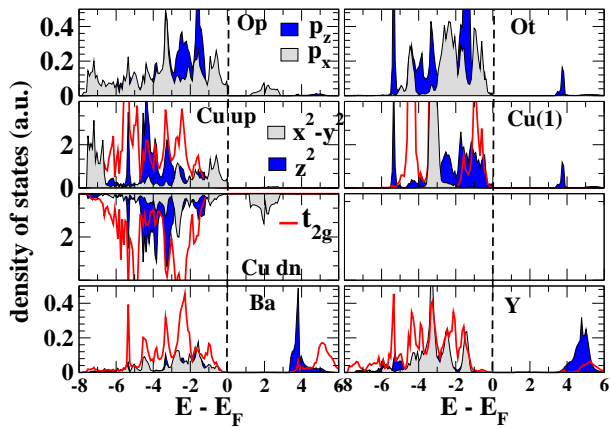}
\caption{OR-DOS of AF YBa$_2$Cu$_3$O$_6$. O$(2)$ and O$(4)$ are in-plane and apical oxygens, respectively. Only the
most important orbitals (d for Cu, Ba, and Y, and
p for O) are shown; solid line is the sum of the 3 nearly degenerate t$_{2g}$ DOS.
For O$_P$ p$_y$ $\sim$p$_z$, while for O$_T$ p$_x$=p$_y$. Supercell: 48-atoms 2$\times$2.
\label{dos_48}}
\end{figure}

\begin{figure}
\includegraphics[clip,width=8cm]{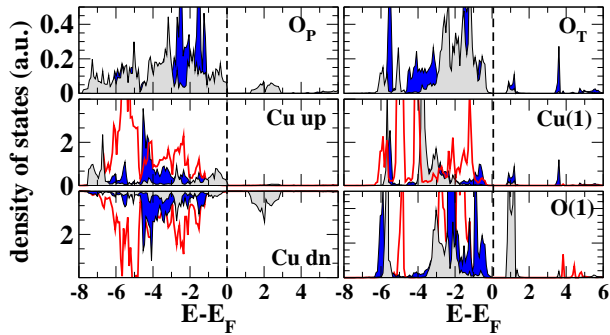}
\caption{OR-DOS of AF YBa$_2$Cu$_3$O$_{6.25}$ for the broken-chain configuration (a) in Fig.\ref{disegno}. Labels are same as Fig.\ref{dos_48}.
\label{dos_49}}
\end{figure}

\begin{figure}
\epsfxsize=6cm
\includegraphics[clip,width=7cm]{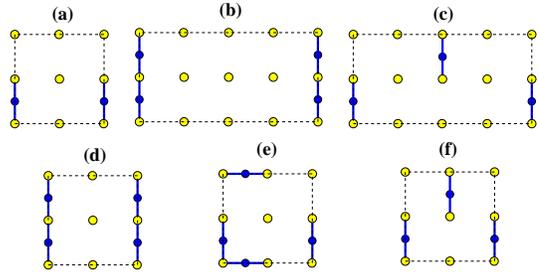}
\caption{(Color online) Doping configurations in the Cu(1)-O chains.
Dark (blue) and light (yellow) circles are O(1) and Cu(1) ions, respectively. 
(a): $x=0.25$, cell 2$\times$2; (b) and (c): $x=0.25$, cell 4$\times$2; (d), (e), and (f): $x=0.5$, cell 2$\times$2.
\label{disegno}}
\end{figure}

The realistic description of underdoped cuprates  is  troublesome for first-principles band theories such as local-spin density (LSDA) or generalized gradient approximation (GGA) density-functional theory \cite{pickett}, which incorrectly describe the  antiferromagnetic Mott insulating state as non-magnetic and metallic, and therefore are not usable to study the IMT. Here we circumvent this issue employing the pseudo-self-interaction correction method (PSIC) \cite{fs}, which is particularly accurate for strongly correlated materials \cite{varie}. As a backup and countercheck, we also use GGA+U \cite{anisimov} and GGA calculations. PSIC calculations are carried out with a plane wave basis ultrasoft pseudopotential \cite{uspp} method, with 30 Ryd cut-off energies. For GGA and GGA+U we employ the PAW method as implemented in the VASP \cite{vasp} code. We use 6$\times$6$\times$6 special k-point grids for total energy calculations, 250 special k-points and linear tetrahedron interpolation method for DOS. The GGA+U parameters are U=10 eV, J=1 eV, giving the best match with PSIC bands. At different doping concentrations, we use supercells  ranging from 24 ($\sqrt{2}$ $\times$$\sqrt{2}$$\times$1 primitive cells) up to 96 atoms (4$\times$2$\times$1). Unless otherwise specified, the results shown below are obtained by PSIC.

In Fig.\ref{dos_48} we show the orbital-resolved density of states (OR-DOS) for tetragonal AF YBa$_2$Cu$_3$O$_6$, which shows a Mott-Hubbard charge-transfer band gap $\sim$1.2 eV (the observed photoconductive threshold at $\sim$1.5 eV \cite{yu}). The valence band top (VBT) and conduction band bottom (CBB) are hybrids of unpolarized O (p$_x$, p$_y$) and, respectively, majority (occupied) and minority (unoccupied) Cu d$_{x^2-y^2}$ states. The fundamental transition only involves orbitals within CuO$_2$ planes; transitions involving  final states with apical O(4) p$_z$ and chain Cu(1) d$_{z^2}$ states (as well as Ba and Y states) start only above $\sim$3.5 eV (more details will be presented elsewhere).

\begin{figure}
\includegraphics[clip,width=8.7cm]{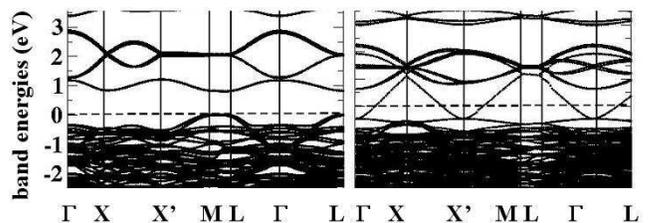}
\caption{Band structure of AF YBa$_2$Cu$_3$O$_{6.25}$. Left: broken-chain (configuration (a) of Fig.\protect\ref{disegno}). Right: full chains (configuration (b), same Figure).}
\label{band}
\end{figure}

Now we consider oxygen inclusion in the chains.  In Fig.\ref{dos_49} we show the OR-DOS at $x$=0.25 calculated with one doping oxygen per 2$\times$2 unit cell (configuration (a) of Fig.\ref{disegno}) corresponding to an alternated array of parallel, isolated Cu(1)-O(1)-Cu(1) segments: this is our broken-chain configuration. The main feature induced by  O(1) is a spin-degenerate hole state $\sim$1 eV above the VBT, strongly localized on O(1) itself with dominant p$_x$ character ($x$ is the Cu(1)-O(1) direction). This hole is residually extended (Fig.\ref{dos_49}) on Cu(1) d$_{x^2-y^2}$
(through dominant pd$\sigma$ coupling) and, more conspicuously, on Cu(1) d$_{z^2}$. This is because d$_{z^2}$ is the highest (most antibonding) Cu(1) d state (see Cu(1) OR-DOS in Fig.\ref{dos_48}). The hole further propagates to apical O(4) p$_z$ states (see Fig.\ref{dos_49}), and up into the CuO$_2$ planes. The calculated  orbital decomposition of the hole band gives 32.3\% on O(1), 33.5\% on the two adjacent Cu(1), 27\% on the apical O(4) closer to the latter, 2.5\% on O(2), 1.3\% on Cu(2), and finally 2.3\% on Ba. The corresponding band structure (left panel of Fig.\ref{band}) shows  the flat hole band running through the whole Brillouin zone just below the undoped conduction band (actually almost touching it at $\Gamma$).

\begin{figure}
\epsfxsize=9cm
\includegraphics[clip,width=8cm]{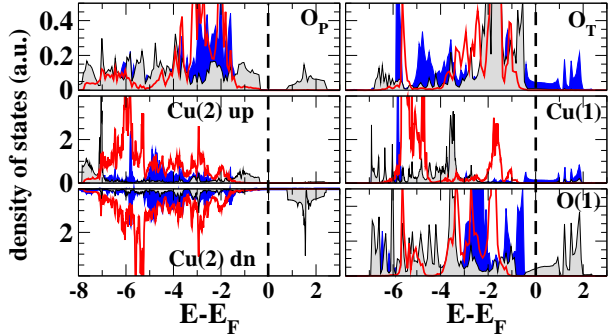}
\caption{Orbital-resolved DOS of ordered-chain AF YBa$_2$Cu$_3$O$_{6.5}$. The calculation is done in  configuration [d] of Fig.\ref{disegno}. Note that the while dispersed band crossing E$_{\rm F}$ has sizable Cu(1)-O(1)-O(4) character, the local DOS in the CuO$_2$ planes is still insulating.}
\label{dos_50}
\end{figure}

To evaluate the effects of oxygen alignment at $x$=0.25 (configuration (b) of Fig.\ref{disegno}) we employed a 4$\times$2 (i.e. 98-atom) supercell, whose band structure is reported in the right panel of Fig.\ref{band}. Now the hole band is dispersed and crosses the Fermi energy, so the system is metallic. The dispersion is one-dimensional: no metallicity is injected into the CuO$_2$ layer, as signaled by the band flatness along the M-L k$_z$ segment. Even at $x$=0.5 the OR-DOS of the chain-ordered AF configuration (Fig.\ref{dos_50}) shows that the hole band does not affect the band gap in the AF CuO$_2$ layer, and spreads instead on the same in-chain orbitals previously seen to be involved in the localized hole decomposition.

So far we have shown that the  order-disorder and one-dimensional IMT in the chains are directly related. This relation holds for   $x$=0.25 and $x$=0.5, so it must hold for any 0.25$<$x$<$0.5. Of course the modeling of disorder in our calculations is limited by the size of our supercell: for a realistic sample of disordered distributions we may envisage a hole energy distribution spanning (and eventually closing) the $\sim$1 eV energy interval above the VBT. So, rather than as an optical insulator, the chains electronic ground state is probably better described as a hopping-conductive Fermi glass, in agreement with the mainstream interpretation. However, the limits in simulating disorder effects do not change the conclusion that a) only chain-alignment produces hole band dispersion, b) the hole band is one-dimensional in character, and the corollary that c) as long as the AF ordering is retained, metallic conductivity does  is not transferred outside the chains. 

We now discuss the energetics of chain-aligned and unaligned configurations at $x$=0.25 and $x$=0.5. The data for both AF and PM ordering, obtained with different approaches, are listed in Table \ref{tab_en}. The chain-ordered configuration is the most stable by a safe
margin, irrespective of CuO$_2$ magnetic ordering. Thus $\Delta$E can be thought as the cost of breaking a single chain moving an O(1) to the nearby isolated segment. These results imply a strong drive towards chain formation and hence one-dimensional metallic behavior. Although disorder does not allow the formation of macroscopic chains at low doping \cite{lee}, our calculated energies confirm the observation that chain nucleation starts at very low doping, also accompanied by local one-dimensional metallic percolation.

\begin{table}[h]
\caption{Total energies for the O(1) configurations in Fig.\ref{disegno} ([b] and [c] for $x$=0.25, [d], [e], and [f] for $x$=0.5), and AF and PM phases (values labeled with * and ** refers to GGA+U and GGA calculations, respectively).}
\label{tab_en}
\centering
\begin{tabular}{lccccc}
\hline\hline
                    &      \multicolumn{2}{c}{$x$=0.25}       &\multicolumn{3}{c}{$x$=0.5}  \\
                    & [b]& \multicolumn{1}{c}{[c]}                               & [d] & [e]   & [f] \\
 \hline
  $\Delta$E$^{PM}$  &  0 & 1.06 (0.80$^*$, 0.69$^{**}$)    & 0   & 0.78 &  1.08 \\
  $\Delta$E$^{AF}$  &  0 & 1.07 (0.98$^*$)                 & 0   & 0.82 &  1.12  \\
\hline\hline
\end{tabular}
\end{table}

\begin{figure}[h]
\centerline{\includegraphics[clip,width=5cm]{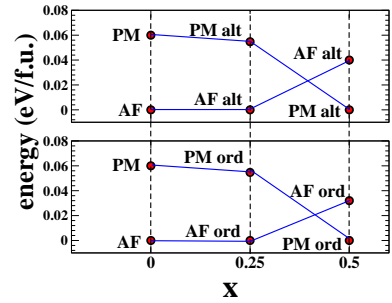}}
\caption{Comparison of the energies of AF and PM phases as function of $x$ for broken-chain (top) and ordered-chain (bottom) configurations.}
\label{energy}
\end{figure}

Next, in Fig.\ref{energy} we  compare the AF and PM total energies for chain-ordered ([b] or [d]) and chain-alternated ([c] or [f]) configurations as a function of $x$. The  energy zero is the lowest phase at each doping. The AF phase is favored up to $x$=0.25 whereas the PM phase is favored at and above $x$=0.5.

\begin{table}
\caption{Differences (in 10$^{-3}$ unit charge) in atomic orbital occupation between $x$=0 and $x$=0.5 for PM and AF ordering and chain configurations (d) and (f) (see Fig.\ref{disegno}). Positive (negative) charges are hole (electron) fractions. The number of equivalent atoms in the cell is in square brackets. Subscript "1" indicates atoms next-neighbors of O(1) or in apical units containing O(1); subscript "2" non-neighbors of, or in units without, O(1): e.g. O(4)$_1$ is first-neighbor to a Cu(1)-O(1) unit; O(2)$_1$ is on-top of O(1), O(2)$_2$ and O(2)$_3$ are co-planar to O(2)$_1$; Cu(1)$_1$ and Cu(1)$_2$ are first- or second-neighbors of O(1); Cu(2)$_1$ and Cu(2)$_2$ are on-top of Cu(1)$_1$ and Cu(1)$_2$, respectively.
\label{tab_occnum}}
\centering
\begin{tabular}{lrrrr}
\hline\hline
	    &   PM$_{\mbox{d}}$   & AF$_{\mbox{d}}$ &  PM$_{\mbox{f}}$ & AF$_{\mbox{f}}$ \\
\hline\hline
doping & & & & \\
\hline
O(1) [2] &  --403    & --401 &  --364  &  --361 \\
total O(1) & --806   & --802 &  --728  &  --722 \\
\hline
  apical  &  &  & &  \\
\hline
 O(4)$_1$ [4] & 65  & 67 & 34 & 35 \\
 O(4)$_2$ [4] & 1 & 11  & 34 & 35  \\
 Cu(1)$_1$ [2] &  189  &  190 & 103  &  105 \\
 Cu(1)$_2$ [2] &  35 &  36 & 103  &  105 \\
 CuO$_2$   [4] & 187  &  191 &   170   &  174 \\
 total Cu(1)   &    448   &  451  & 413  & 418 \\
 total Cu(1)+O(4) & 747  &  764  & 682 &  697 \\
\hline
 planar & & & & \\
 \hline
 O(2)$_1$  [4] & 3  & 10   & 2  & 7 \\
 O(2)$_2$  [4] &--2  & 3  & --2 &  3 \\
 O(2)$_3$  [8] & 2   & 7 & 1  & 5 \\
 total O(2)    & 23  & 106 & 10 &  82 \\
 Cu(2)$_1$ [4] & 13   & --4& 10 &--4  \\
 Cu(2)$_2$ [4] & 12  & --5& 1 & 4 \\
 CuO$_2$ [8]   & 15  &  9 & 11 & 6\\
 total Cu(2)   & 98  &  --34 &  77 & --34 \\
 total Cu(2)+O(2) &  121  &   72 &  86 & 48 \\
 \hline 
  Y [4]        & --3 & 0  & --3 & 0 \\
  Ba [8]       & --4 & --1  & --2 &   0    \\
\hline\hline
\end{tabular}
\end{table}

Three points are  worth noticing. First, the  AF$\rightarrow$PM crossover (with attendant in-plane insulating$\rightarrow$metallic transition)  occurs in the doping interval where superconductivity sets on. Second, the AF-PM competition is barely affected by the O(1) distribution within the chains; this implies in both instances a weak chain-plane coupling with  minimal hole transfer to the planes, confirming the high inefficiency of  oxygen doping. Third, the AF-PM competion is quite tight (a few tenths of meV per formula), in line with  the tendency of the AF phase to form cluster spin-glass states and the coexistence of microscopic AF and PM domains \cite{niedermeyer}. 

The AF-PM competition can be interpreted analyzing the changes in orbital occupation upon doping, reported in Table  \ref{tab_occnum}. Each O(1) takes over $\sim$0.4 electrons. This quantity is quite independent of  magnetic ordering, but is visibly larger for configuration (d) than for (f): full-chain order enhances charge transfer between the doping O and its surrounding. The vast majority of the hole released by O(1) remains in the apical units (Cu(1)-O(1)-O(4)), and only a fraction is transferred to the planes. The most receptive environment is PM$_{\rm d}$, which collects 13.9\% of the hole charge into the planes, followed by PM$_{\rm f}$ (11.2\%), AF$_{\rm d}$ (8.7\%) and AF$_{\rm f}$ (6.4\%). 

We can understand this classification in terms of magnetic and chain ordering. The PM phases have in-plane CuO$_2$ bands at E$_F$, 
allowing hole injection in the highly receptive Cu(2) atoms. In AF phases the Cu(2) states are swept off the E$_F$ region and are marginally affected by hole injection. Table \ref{tab_occnum} shows that the PM  in-plane hole population is mainly on  Cu(2), and in the AF phase in-holes mainly resides on O(2), with  Cu(2) having small electron-like excess charge. As for O(1) configuration, (d) ordering is a better environments for hole transfer than (f). The metallic band originating from chain ordering, with its hybridization with apical atoms, is better suited to charge transfer than insulating disordered states. 

Finally, the charge transfer efficiency is the key to understand the PM-AF competition. Charge injection in the planes tends to stabilize the PM phase at increasing doping. At $x$=0.5, the AF-to-PM transition is accompanied by  increase of $\simeq$0.05 holes per CuO$_2$ unit for both the (d) or (f) configurations, i.e. the AF-PM hole charge imbalance does not depend on the O(1) configuration. This explain the substantial similarity of the two panels in Fig.\ref{energy}.

In summary, we described from first principles the effects of oxygen doping in underdoped YBa$_2$Cu$_3$O$_{6+x}$. We found chain alignment energetically favored with respect to broken chains at any doping concentration. Since oxygen alignment ordering produces one-dimensional metallic bands, chain metallicity is formally present at any doping, although over short distances only. We also considered the energy competition of the insulating AF and metallic PM state in the CuO$_2$ planes, showing that the latter phase is favored  above a threshold doping located between $x$=0.25 and $x$=0.5. The in-plane AF-PM transition is caused by the (sparing) chain-to-plane hole transfer, but is quite independent of the oxygen ordering in the chains. 

Work supported in part by MiUR  through projects ``Cervelli per la ricerca'', PON-Cybersar, PRIN 2005, and by Fondazione Banco di Sardegna (Project Correlated oxides).

\end{document}